\documentclass[nofootinbib,aps,onecolumn,letterpaper,preprintnumbers,superscriptaddress,showpacs]{revtex4}
\usepackage{amsmath}
\usepackage{eurosym}
\usepackage{amssymb}
\pdfoutput=1
\usepackage{graphicx}
\usepackage{color}
\usepackage{braket}
\usepackage{amsmath}
\usepackage{amssymb}
\usepackage{subfigure}
\usepackage{amsfonts}
\usepackage[left=1.5cm,right=1.5cm,top=1.5cm,bottom=1.5cm]{geometry}
\usepackage[usenames,dvipsnames,svgnames]{xcolor}  %% colored text
\usepackage{hyperref}   %% Needs to make hyper reference of any references or citations
\definecolor{beamer@PRD}{RGB}{46,48,146}

\begin{document}
\newcommand\be{\begin{equation}}
\newcommand\ee{\end{equation}}
\newcommand\bea{\begin{eqnarray}}
\newcommand\eea{\end{eqnarray}}
\newcommand\bseq{\begin{subequations}} %solo con amsmath
\newcommand\eseq{\end{subequations}}
\newcommand\bcas{\begin{cases}}
\newcommand\ecas{\end{cases}}
\newcommand{\p}{\partial}
\newcommand{\f}{\frac}

\title{Probing Planck Scale Spacetime By Cavity Opto-Atomic $^{87}$Rb Interferometry}

\author {Mohsen Khodadi}
\email{m.khodadi@stu.umz.ac.ir}
\affiliation{Department of Physics, Faculty of Basic Sciences,\\
University of Mazandaran, P. O. Box 47416-95447, Babolsar, Iran}

\author {Kourosh Nozari}
\email{knozari@umz.ac.ir (Corresponding  Author)}
\affiliation{Department of Physics, Faculty of Basic Sciences,\\
University of Mazandaran, P. O. Box 47416-95447, Babolsar, Iran}
\affiliation{Research Institute for Astronomy and Astrophysics of Maragha
(RIAAM), P. O. Box 55134-441, Maragha, Iran}

\author{Anha Bhat}
\email{anhajan1@gmail.com}
\affiliation{Department of Metallurgical and Materials Engineering, National Institute of \protect \\Technology, Srinagar 190006, India}

\author{Sina Mohsenian}
\email{sina$_$mohsenian1@uml.edu}
\affiliation{Department of Mechanical Engineering, University of Massachusetts Lowell, Lowell, Ma, USA}

 %---------------------------------------

\begin{abstract}
The project of \emph{``quantum spacetime phenomenology''} focuses on searching pragmatically for the Planck
scale quantum features of spacetime. Among these features is the existence of a characteristic
length scale addressed commonly by effective approaches to quantum gravity (QG).
This characteristic length scale could be realized, for instance and simply, by generalizing the standard Heisenberg uncertainty principle
(HUP) to a \emph{``generalized uncertainty principle''} (GUP). While usually it is expected that phenomena belonging to the realm of QG are essentially probable
solely at the so-called Planck energy, here we show how a
GUP proposal containing the most general modification of
coordinate representation of the momentum operator could be probed by a
\emph{``cold atomic ensemble recoil experiment''} (CARE)
as a low energy quantum system. This proposed atomic interferometer setup has advantages over the conventional
architectures owing to the enclosure in a high finesse optical cavity which is supported by a new class of low
power consumption integrated devices known as \emph{``micro-electro-opto-mechanical systems''} (MEOMS).
%The proposed system comprises of a micro mechanical oscillator instead of spherical
%confocal mirrors as one of the components of high finesse optical cavity.
In the framework of a top-down inspired bottom-up QG phenomenological
viewpoint and by taking into account the measurement accuracy realized
for the fine structure constant (FSC) from the Rubidium ($^{87}$Rb)
CARE, we set some constraints as upper bounds on the characteristic
parameters of the underlying GUP. In the case of superposition of the
possible GUP modification terms, we managed to set a tight constraint
as $0.999978<\lambda_0<1.00002$ for the dimensionless characteristic
parameter. Our study shows that the best playground to test QG approaches
is not merely the high energy physics, but a table-top nanosystem assembly,
as well.
\end{abstract}
\maketitle

\section{Introduction}
Despite three well-tested and powerful theories: quantum field theory (QFT),
standard model of particles (SM) and general relativity (GR), our theoretical
framework for studying the universe is incomplete for two main reasons.
The primary reason is that no well-formulated quantum gravity (QG)
theory that unifies the micro level universe (subjected to quantum mechanics (QM))
and Einstein’s theory of gravity (relevant to macro level
universe) has been constructed properly so far. The secondary reason is the lack of a
well-established framework for unification of gravity with three
other fundamental interactions. The root of this incomplete issue comes back to
substantive differences between QM and GR, which are well discussed in, for
instance, Refs.~\cite{H2003, Hedrich2009}. Nowadays, there is a great curiosity
in theoretical physics community about understanding the physics of the Planck scale
which is considered to be one of the challenging research areas. Given that QM
quantizes any dynamical field as well as its physical content, so QG would
be related to the discreteness of geometrical concepts such as, length, area
and volume. This implies that the combination of the relativistic and quantum
effects results in reshaping the common thought of distance around the natural
scale known as \emph{''Planck scale''} $\ell_{pl}\approx10^{-35}$ m. Although,
a fully functional unified QG theory has remained out of reach for less than a
century, however there are some convincing approaches available such as string
theory \cite{st1,st2}, canonical quantum gravity \cite{Lee, Ashtekar}, various models of deformed
special relativity (DSR) \cite{dsr1,dsr2}, and also a variety of generalized uncertainty principle
(GUP) \cite{Ali2009, Das2010}, which seem to be potential candidates for converging the study.
These approaches implicitly predict a fundamentally quantized space due to the existence of a minimal
uncertainty in physical distance $\Delta x_{min}$ of the order of the Planck
length $\ell_{pl}$. In fact Einstein's perspective on gravity as a property of spacetime, and not
merely as a usual force, has led to make an effective framework of QG with the impression
that a quantum particle is moving on a spacetime with a geometry equipped with a fundamental, minimal characteristic length. It can be shown easily that the
minimal length uncertainty results in a modification of the standard canonical
commutation relations (CCR) between position and momentum and can be interpreted as a
Lorentz invariant natural cutoff, \cite{Kempf1}. In other words, the concept of minimal
physical length appears naturally in QG scenarios by generalization of the ordinary
Heisenberg commutator relations between position and momentum in Hilbert space. In this respect,
it is important to note that all the above mentioned approaches to QG can be considered as
the origin of GUP which allow the study of the effects of a minimal length scale in different
areas of physics,  see \cite{Sabin2012} for a detailed review.
So, the deformation of Heisenberg's uncertainty principle (HUP), that is, GUP,
is a natural and inevitable consequence of QG proposal. As an additional
motivation for emphasizing the theoretical importance of GUP, let's to mention attempts for resolution of the
information loss puzzle in black hole physics by considering GUP, see for instance \cite{Kempf1, Kempf2, Kempf3, Amati, M1, M2, M3}.
Concerning the black hole physics, in \cite{Adler2001} argued that there could be black hole remnant due to GUP.
In this direction, the implication of GUP on the complementarity principle proposed for the
black hole information loss paradox, investigated in \cite{Chen2014}.
Also in \cite{Black2012} it is shown that by taking GUP in the context of extra dimensions theories,
there is a possibility to agree with the outcomes of LHC which claims no black holes
are recognizable around $5$ TeV \cite{CMS2011}. Since a deformation of Heisenberg uncertainty
principle affects various quantum phenomena, some efforts have been made to understand phenomenological aspects of GUP models in the context of astrophysics, cosmology and high energy physics in recent years, see for instance \cite{Sabri, ref:PLB, ref:JCAP, kh2008, Tawfik,ref:PLB2016, ref:GRG, ref:Ann, ref:CQG}. The consequence of GUP on the physics of white dwarf stars also studied in \cite{Reza2016, Yen2018, Kumar2018}.
Violation of weak equivalence principle within QM is one of the controversial outcomes of GUP \cite{WEP1, WEP2} which the
deviation reported in neutron interferometry experiment \cite{Interf2015} confirms it.\\

The lack of a direct experimental test of tentative theories of QG
is due to the limits, both in measurement and huge energy required in some controllable experimentally phenomena in
the vicinity of the Planck scale. However, this attempt has evolved seriously as one of the major
challenges of the QG phenomenological studies in recent years, see for instance Refs.\cite{Amelino, Sabin}.
In the light of proposed experimental tests, the current path of QG could
be more coherent since naturally it is expected that some of the involved proposals
may be redundant and even ruled out. Indeed, it is expected that application of some
experimental practicality within the different proposals related to QG
make them more efficient and realizable depending on the engineered
improvements in the current designs. Within the last two decades, technical developments in experiments related to phenomenology of
quantum gravity and probing spacetime structure at short distances were focused on the control of
physical systems at the Planck scale. These attempts were able to provide
promising facilities in order to check correctness of the phenomenological predictions
of QG theories \cite{Stachel}. The already conceived idea of QG tempts one to think that the intersection between QM
and GR is too difficult to be accessed through laboratory experiments and seems
indeed to be on the other side of the accessible knowledge horizon \cite{Lect2000}. In other words,
it was assumed that firm predictions are impossible without a fully fledged
theory of QG, which combines QM and GR in a new set of laws comparable with observations. However, in a new look at QG phenomenology,
that has started with the works referred in Refs.~\cite{G1998, DV1998, G1999, GF1999},
it is believed that there is the possibility of meeting the theory with
experiment even in the absence of a well-established, final theory of QG.
The new aspect of QG phenomenology can be defensible from the following two perspectives.
Firstly, over the time the empirical techniques have made significant progress that has
now made it possible to test the QG effects credibly. Secondly, theoreticians succeeded in
handling some proposals and ideas with permissible prospects that could be tested by common
experimental techniques. In a generic sense, the foremost purpose of QG phenomenological
studies is to try to suggest some ways of deriving potential experimental signatures
from numerous approaches. The modifications arising from QG in the form
of relevant GUP(s) affect Hamiltonian of systems under study, expected to lead to small
modifications in measurable quantities in the bed of some physical experiments.
So far, numerous attempts have been made to explore spacetime structure at the Planck scale by
tuning the characteristic parameter introduced by the relevant QG proposal with
experimental data, see Refs.~\cite{UP0, UP1, UP2, UP3, UP4, UP5, UP6, UP7, UP8}.
The upper bounds released so far strongly indicate the fact that one needs to look for even more advanced
and special experimental setups to detect QG effects with ideal resolution expected in
theory. Of course, the progress made in recent years in coherence between
theoretical and experimental aspects of quantum optics, in particular the new approach towards
light-matter interactions, has opened a novel direction for testing new areas of physics
such as Planck scale physics with excellent resolution \cite{Nature, NPB, Sci}.\\

One of the most accurate and trustworthy tools that can be used to explore new areas
of physics is \emph{``atomic interferometry''} (AI) which in its advanced setups is exploited
\emph{``laser cooled atoms"}. It is important to note that because
of certain sensitivity of AI, it has  been used extensively to achieve accurate measurements
in physics \footnote{AI systems also have some other applications like that of being developed as accelerometers \cite{Accel}
and gyroscopes \cite{Gyro}. It is interesting to mention that recently such systems were used to map the gravitomagnetic
field of the earth in a circling satellite \cite{Map}} \cite{Setup2006, PRA2016, RMP2009}.
Also in recent years we faced with impressive application of AI for testing
fundamental laws and principles of physics \cite{Interf2015, PRL2018}. Amelino-Camelia et.al were pioneer to propose the idea that
AI \cite{Atomic2002} can be used for controlling the QG characteristic parameter released in
DSR deformed dispersion relation \cite{Camelia2009}. This precious idea has motivated
us in this paper to investigate the possibility of probing the Planck scale
spacetime with resolution allowed by the Rubidium cold atom embedded in an AI
setup, proposed based on a new class of microsystems technology known as
\emph{``Micro-opto-electro-mechanical systems''} (MOEMS)\footnote{In a simple term,
MOEMS is a subset of microsystems technology which together with Micro-electro-mechanical
systems (MEMS) forms the specialized technology fields using miniaturized combinations
of optics, electronics and mechanics.}. As one of the most
sensitive and accurate measurements performed by AI that plays
a central role in our analysis is determination
of the \emph{``fine structure constant''} (FSC) $\alpha$. In more details, AI has managed
to obtain a value of $\alpha^{-1}=137.035999037(91)$ \cite{Alpha} which is the second
best value relative to the one concluded from the electron anomaly measurement\cite{Electron}.
This has prompted the investigation that AI can provide the possibility of exploring Planck scale spacetime with
improved resolution. In this manuscript, we confront a GUP proposal which contains the most
general deformation as proposed in \cite{Mir2016} with cold atom recoil measurement.
The proposed study predicts the narrowest constraints, considering all modification terms
up to the highest possible order of GUP characteristic scale.

\section{Natural Cutoffs Modified Commutation Relations: the Most General Form of Modification}

Common versions of GUP comprise of modification terms containing linear and quadratic
momentum operator or their combinations  \cite{Ali2009, Das2010}. Here, we plan to introduce
a GUP proposal containing a general deformation function which is expected to
address the most general form of modification of momentum operator. Our idea
explicitly is that by taking higher powers of momentum in its deformed coordinate
representation, there may be phenomenologically the possibility of probing Planck
scale spacetime with an improved resolution.\\

The most general form of GUP can be suggested through the following commutation relations
\begin{eqnarray} \label{e-0}
[x^i, p_j] = i \hbar \left(\delta^i_j + f[p]^i_j\right)~,
\end{eqnarray} in which $f[p]^i_j$ denotes a tensorial function so that its functional
dependency to momentum strongly depends on the form of deformed coordinate representation
of the momentum operator admitted by various GUP models \cite{Mir2016}.
By employing an inductive method one can obtain such expression as
\begin{eqnarray}\label{e-1}
 p_i  \to \tilde p_i =  p_i \left(1 + \lambda_0+\lambda_1 (p^j p_j)^{1/2}+
 \lambda_2 (p^j p_j)+\lambda_3 (p^j p_j)^{3/2}+\lambda_4 (p^j p_j)^{2}+.... \right)~,
\end{eqnarray} for the most general form of the modified momentum operator.
The above expression can also be demonstrated in a more compact form as follows
\begin{eqnarray}\label{e-2}
 p_i  \to \tilde p_i =  p_i \left(1 +  \sum  \lambda_{n} (p^j p_j)^{n/2} \right),~~~~
 n=0,1,2,3,.....
\end{eqnarray} in which $\lambda_{n}=\frac{\lambda_0}{(M_{P}c)^n}$
is the most general form of GUP characteristic parameter containing dimensionless
constant $\lambda_0$. Of course, if we insist on the condition $\lambda_0>0$ to be satisfied,
the most general form of the modified momentum operator (\ref{e-2}), re-expresses as
\begin{eqnarray}\label{e-3}
 p_i  \to \tilde p_i =  p_i \left(1 \pm \sum  \lambda_{n} (p^j p_j)^{n/2} \right),~~~~
 n=0,1,2,3,.....
\end{eqnarray} where in which follows we focus on this version. Here, should be
stress that although in the presentation (\ref{e-2}), both cases $\lambda_0>0$ and $\lambda<0$
are possible but due to the form of (\ref{e-3}), we deal with only with positive values.
Although from the perspective of phenomenology, it is believed that the GUP parameter
should be positive, however in \cite{Yen2018,Ye2018} the possibility of a negative\footnote{According to literatures the
possibility of negative sign for GUP parameter was first raised in \cite{Jizba2010} to formulate it on a crystal lattice.} one also discussed.
Theoretically, exploring quantum spacetime cannot be fully achieved
unless in the case of $\lambda_0=1$. This brings the notion that on deriving upper
bounds close to unity, the quantum spacetime becomes more accessible which
is a step forward in the direction of testing relevant QG predictions.
In Ref.~\cite{Mir2016}, further details can be found on the GUP
proposal at hand.
\begin{figure}[ht]
\includegraphics[width=14cm]{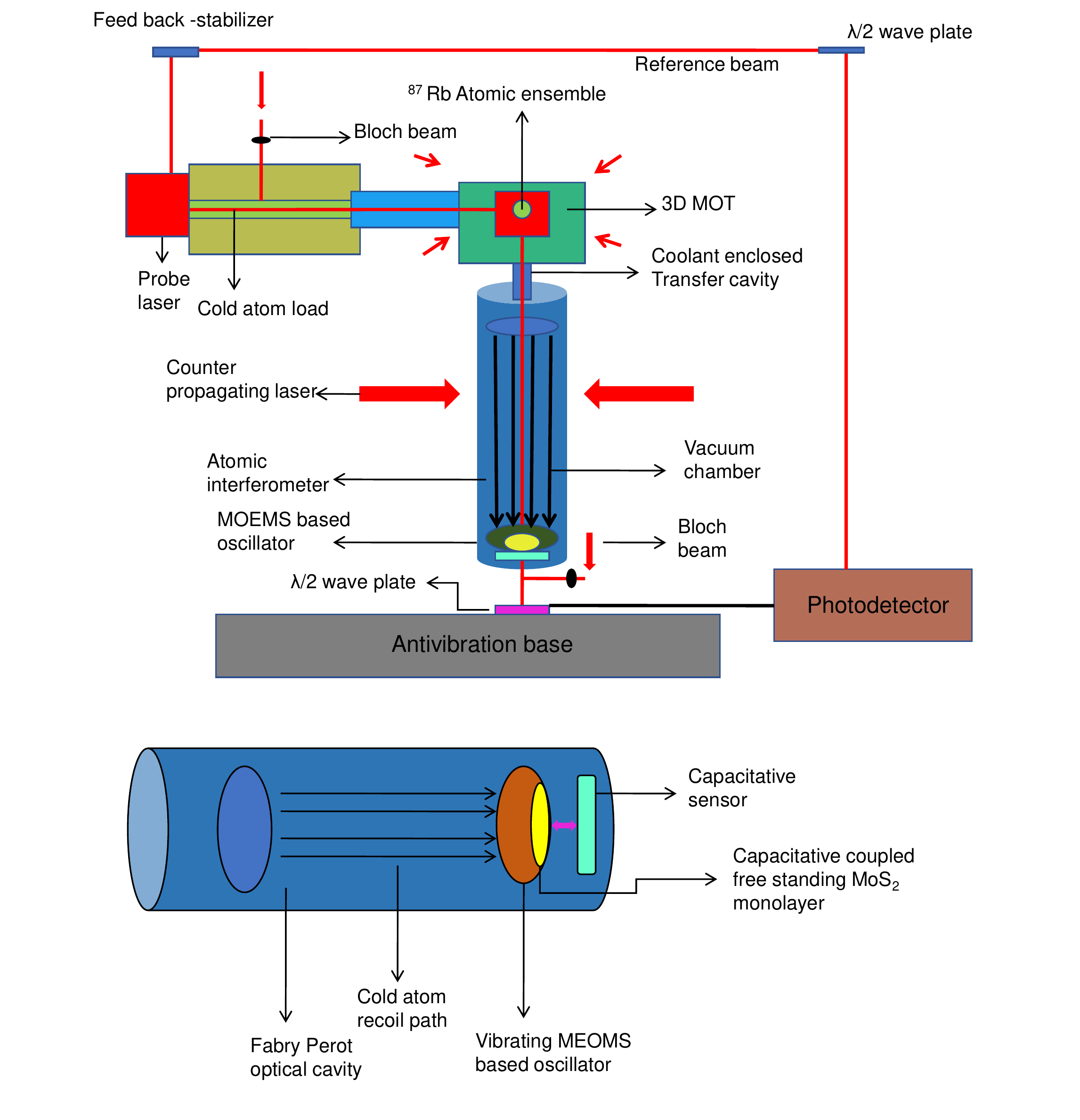}
%\hspace{0.5cm}
%\includegraphics[width=6cm]{f1b.pdf}
\hspace{0.5cm}
\caption{Schematic representation of inbuilt atom interferometer MEOMS based optical finesse cavity.}
\label{Fig1}
\end{figure}

\section{Experimental Scheme: Inbuilt atom interferometer MEOMS based optical finesse cavity }

\begin{figure}[ht]
\includegraphics[width=10cm]{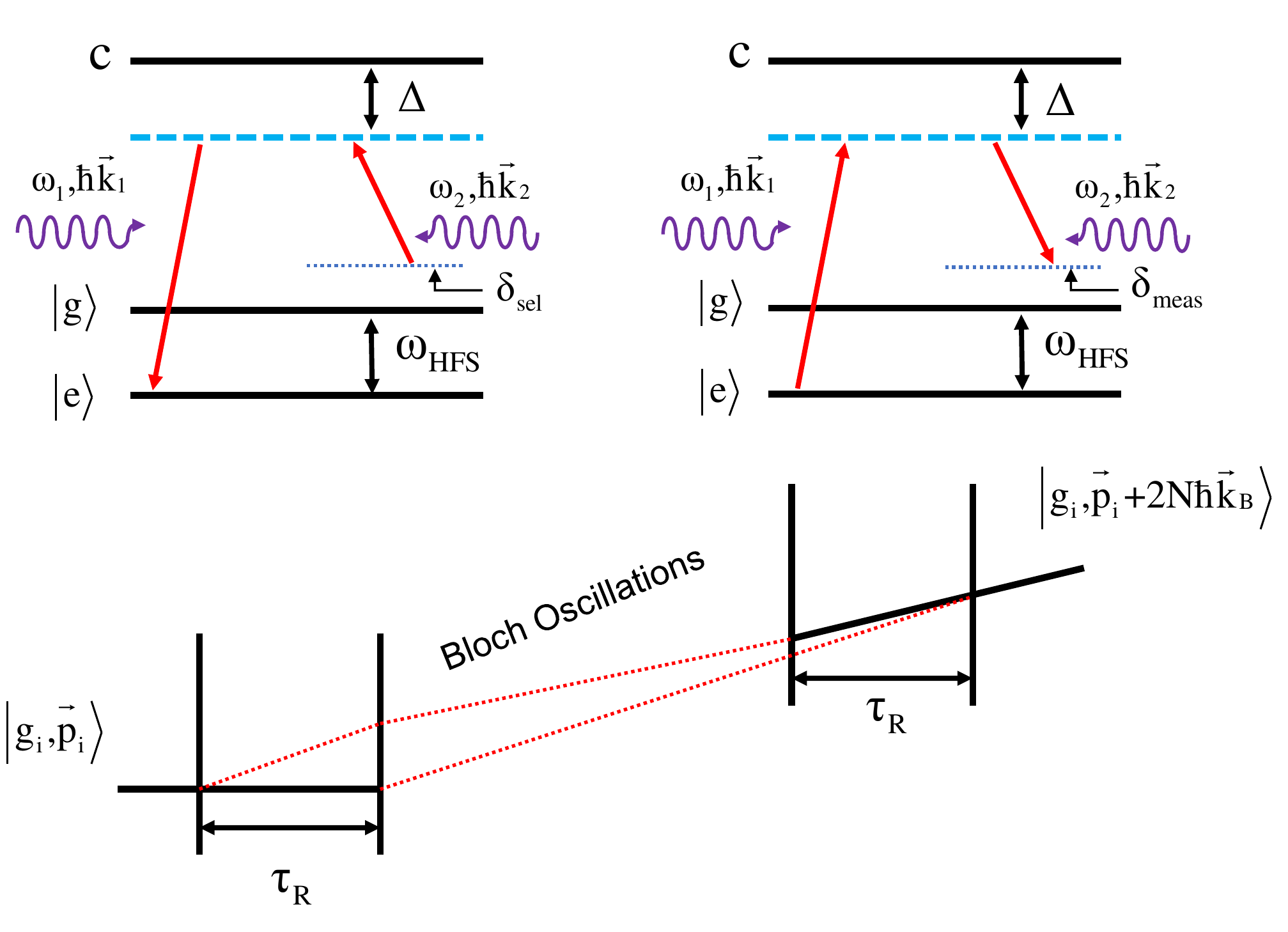}
%\hspace{0.5cm}
%\includegraphics[width=6cm]{f1b.pdf}
%\hspace{0.2cm}
\caption{Schematic representation of Raman transition
between hyperfine levels $|g\rangle$ and $|e\rangle$
along with Bloch oscillations.}
\label{Fig2}
\end{figure}

The basic principle of this assembly is that the cold ensemble of
atoms, when is launched inside an optical cavity, splits into two clouds
which both traverse different paths based on the laser input. This
generates a velocity-based selection of atoms based on their excitation
in the hyperfine states, allowing them to recoil backwards or be defined still by the states $|g\rangle$
and $|e\rangle$. The difference between recoiling atoms and steady atoms
sets a number of Bloch oscillations which transmits the recoiling velocities
to the slowed atoms and ensures that the internal state of atom does not change.
The atomic velocity increases by $2u_r$ where $u_r$ is the recoil velocity of
the atom when absorbing the momentum of the photon. The final velocity at this
stage is the sum of the initial velocity distribution and the $N$ Bloch
oscillations. Meanwhile, another Raman pulse is applied to the final
velocity transferring the atoms back to the $|g\rangle$.
The operation of the proposed system comprises of the engineered Fabry-Perot optical finesse
cavity which conventionally constitutes of two confocal mirrors supporting the
free space propagation of light, while in this design one of the confocal mirrors is replaced by vibrating micromechanical oscillator.
The role of optical finesse cavity has been broadly discussed in several
proposed experiments for probing non-commutative theories and
the possible realization of QG phenomena using the current
technology. The design of optical cavity coupled with nano and micro-mechanical elements have been reported in
previous studies, see Refs.~\cite{NPB, Sci}. However, in the present study we propose a novel confirmation of the
architecture based on the same optical coupling with electromechanical systems, but now
by introducing the atomic ensemble in the cavity and replacing one of the spherical confocal mirrors with micro
mechanical oscillator.
We propose the introduction of MEOMS component which besides electro-mechanical coupling
provides for the integration of optical component as well
for exploring the QG phenomena. This setup is schematically displayed in
Fig. \ref{Fig1}. The Fabry-Perot optical cavity amplifies the
light on each pass which creates a laser gain medium in the
cavity and limits the frequency noises adding to its sensitivity
towards wavelength of the light. The resonant enhancement in
the cavity, increases the transitions due to minimum laser power
which otherwise consumes a multi watt power system. The AI enclosed
in the optical cavity limits the wavefront distortions found in
conventional system which increases its sensitivity and efficiency.
One of the pivotal components in the assembly generally described in
Fig. \ref{Fig1} is a miniature MEOMS based oscillator which also acts
as a complimentary confocal mirror for the cavity and has been designed
in such a way that it acts as a transducing element to pick up the optical
signal for the atomic ensemble. The oscillator keeps moving so as to arrest
the ensemble during its \emph{``Raman transition''} between the hyperfine
levels. The transition should be coherent to the Bloch's oscillation provided
by the probe laser creating about $N$ Bloch oscillations in each run and the
velocity of the atoms could be effectively measured. The periodic
motion of the vibrational MEOMS based oscillator is capable of
limiting the spatial variations to reduce the interferometric
contrast. The MEOMS based oscillator is proposed to be made
of a free standing monolayer of Molybdenum disulphide (MoS$_{2}$)
which is known to have an optical bandgap of about $1.6- 2$ eV
and shows strong photoluminescence near band gap due to
circularly polarized optical pumping. The changes due to
atom recoil on the monolayer MoS$_{2}$, are detected by the low
power consumption photodetector and causes minimum scattering
and loss \cite{G2012}. In this design, the oscillator is attached
to the photodetector which monitors the recoil velocity of the
atomic ensemble without affecting the internal state of the atoms.
The initial leg of the experiment starts with a 2D-MOT (two dimensional-magneto-optical trap)
molasses which is expected to load the atoms into the system
by slow atomic beam into hyperfine level $|g\rangle$. Then using
a standard 3D-MOT molasses it is brought into the vicinity of the optical cavity.
The advantage of this architecture is that it allows the
combination of a high atomic flux and controlled atomic velocities
which were not precisely provided by conventional use of Zeeman
slower or thermal beam devices. As a result, when the cold atoms
are pumped into the cavity by the 3D molasses activation, it provides
a precise control of the velocity in the optical velocity \cite{Muller2009}.
The velocity adjustments contribute to the flat parabolic atomic
trajectories compensating for the default gravitational acceleration
effect. This velocity selection of atoms in the cavity provides for
more flexible time dependent studies like Raman transitions between
two hyperfine ground states and a complimentary Bloch oscillation which is
schematically displayed in Fig. \ref{Fig2}. It is worthy of mention that in this setup the
time dependent noise sources also will limit making it potentially more efficient.
The atomic source is the Rubidium $^{87}$Rb and the ensemble is  induced into
the optical cavity through the MOT arrangements. The atoms exist in between two
hyperfine ground states $|g\rangle$ and $|e\rangle$. The atomic ensemble
in the optical cavity detects maximum possible recoils due to Raman
transition when two counter propagating laser beams excite the atoms that are
trapped in it after the cooling. The Raman frequencies of $\omega_1$ and
$\omega_2$ and wave vectors $\vec{k}_1$ and $\vec{k}_2$ create coherent atomic beams
in the $|e\rangle$ state where the remaining atoms of $|g\rangle$ state
are tuned to single photon transition. The reception of these
transitions in hyperfine states give rise to the well defined $N$ Bloch
oscillations which are well received by the MEOM oscillator.

\subsection{Formulation}

Now we formulated the above description mathematically. The energy conservation law imposes the
following equality at resonance
\begin{equation}\label{conserv}
\hbar (\omega_1 - \omega_2 -\omega_{HFS})= \hbar\Delta + \frac{[\hbar
(\vec{k}_1-\vec{k}_2)+m \vec{u}_i]^2-m^2 \vec{u}_i^2}{2 m}\,.
\end{equation}
Here, quantities $m,~\Delta$ and $ \hbar\omega_{HFS}=E_{g_1}-E_{g_2}$
refer to the mass of atoms, the single photon setting of the atomic
levels and the energy difference between the hyperfine levels
$|g\rangle,~|e\rangle$, respectively. By defining
$\delta_{sel}\equiv \omega_1 - \omega_2 -\omega_{HFS}$
as a Raman transition setting, the above conservation equation
takes the following form
\begin{equation}\label{select}
 \delta_{sel}=\Delta + \frac{\hbar}{2 m}(\vec{k}_1-\vec{k}_2)^2+
 (\vec{k}_1-\vec{k}_2)\cdot \vec{u}_i\,.
\end{equation}

In order to transfer the high recoil velocities to atoms in a short time,
so that the relevant hyperfine internal state of atoms does not change
after process of velocity selection, number of $N$ Bloch
oscillations comes into display. Every Bloch oscillations separately increase the atomic
velocity by factor of 2$\vec{u}_r$, where $\vec{u}_r=\frac{\hbar \vec{k}_B}{m}$
refers to the recoil velocity of the atom while absorbing a photon with relevant
momentum $\hbar \vec{k}_B$. As a result, the final velocity  attributed to the atoms can be denoted by
$\vec{u}_f=\vec{u}_i + 2 N \vec{u}_r$. In the second pair of Raman $\frac{\pi}{2}$ pulses, the
final velocity distribution is centered on $\vec{u}_f$.
Here also by applying the energy conservation law, the Raman regulating
for the velocity measurement reads off as
\begin{equation} \label{meas}
\delta_{meas}=\Delta + \frac{\hbar}{2 m}(\vec{k}_1-\vec{k}_2)^2+(\vec{k}_1-\vec{k}_2)\cdot \vec{u}_f\,,
\end{equation}
where by subtracting it from Eq. (\ref{select}), one gets
\begin{equation}\label{select-means}
|\delta_{sel}-\delta_{meas}|= (k_1+k_2)|u_f-u_i|.
\end{equation}
Now the ratio of Planck constant to the mass of the atoms i.e. $\frac{h}{m}$,
is found by
\begin{equation}\label{h/m}
\frac{h}{m}=\frac{2 \pi |\delta_{sel}-\delta_{meas}|}{2 N k_B(k_1+k_2)},
\end{equation}
which addresses the FSC $\alpha$ via the following well known relation
\begin{equation}\label{alpha}
\alpha^2=\frac{2 R_{\infty}}{c}\frac{m}{m_e}\frac{h}{m}~.
\end{equation}
Here, $R_{\infty},~m_e$ and also $c$ are the Rydberg constant, the electron
mass and the speed of light, respectively. In Ref.~\cite{Alpha}, by the consideration of
the Rubidium $^{87}$Rb atoms the authors performed an excellent measurement of $\frac{\hbar}{m}$
which after an approximate margin of error, has released a value as $\frac{h}{m_{Rb}}=
4.5913592729(57)\times 10^{-9} m^2 s^{-1}$. In this measurement the role of the internal
hyperfine levels $|g\rangle$ and $|e\rangle$ lies in 5S$_{\frac{1}{2}}$ $|$F=2, m$_F$=0$\rangle$
and 5S$_{\frac{1}{2}}$ $|$F=1, m$_F$=0$\rangle$ states of $^{87}$Rb atoms, respectively.
As a necessary information in this measurement, an atomic beam with an initial velocity
$u_i=20$ m/s, was prepared in the $F=2$ hyperfine level. The Ti sapphire laser with
relevant wavelength $\lambda_B=\frac{2\pi}{k_B}=532$ nm was produced  which could produce $N=500$
Bloch oscillations for the application on $^{87}$Rb atoms in each run.

\section{Explicit constraints for Planck scale characteristic parameter}

\subsection{The case without superposition of the GUP modified terms}
Here we address the GUP modifications in terms of different
values of $n$ separately. In a lucid term, we assume that there is
no superposition of GUP modification term which perceives each GUP
modification term to be considered separately in the absence of other
possible terms.\\

By setting $n=0$, Eq. (\ref{e-3}) reads off as
\begin{eqnarray}\label{e-A0}
 p_i  \to \tilde p_i =  p_i (1 \pm \lambda_0)~.
\end{eqnarray}
In this new representation, Hamiltonian related to a one-dimensional quantum
system can be expanded as
\begin{equation} \label{e-A1}
H_{GUP(n=0)} = (1+\lambda_0^2\pm 2\lambda_0)\frac{p^2}{2 m} + V(x)~.
\end{equation}
By applying the above GUP-modified Hamiltonian to the cold $^{87}$Rb atoms,
the relevant GUP term modifies the kinetic energy of $^{87}$Rb atoms as follows
\begin{equation}\label{e-A2}
E_{K-GUP(n=0)}=(1+\lambda_0^2\pm 2\lambda_0)\frac{p^2}{2 m_{Rb}}
\end{equation}
According to the prior pattern, here the GUP counterpart of Eq. (\ref{select-means})
takes the following form
\begin{equation}\label{e-A3}
|\delta_{sel}-\delta_{meas}|_{GUP(n=0)}=(1\pm\lambda_0)^2 (k_1+k_2)|u_f-u_i|~,
\end{equation} where by putting it into Eqs. (\ref{h/m}) and (\ref{alpha})
finally we will arrive at
\begin{equation}\label{e-A4}
\alpha^{-1}_{GUP(n=0)}=\bigg(\frac{2R_\infty}{c}.\frac{m_u}{m_e}.\frac{h}{m_{Rb}}
(1\pm\lambda_0)^2\bigg)^{-\frac{1}{2}}\,.
\end{equation}
Given that the setup at hand is a low energy quantum system, by estimating the
GUP effect on the FSC via the following relation, one can derive an explicit upper
bound for $\lambda_0$
\begin{equation}\label{e-A5}
\frac{|\alpha^{-1}_{GUP}-\alpha^{-1}|}{\alpha^{-1}}<\mbox{measurement accuracy of
$\alpha^{-1}$}
\end{equation}
This relation expresses the fact that by taking into the account the measurement
of $\alpha^{-1}$
with a precision order of magnitude of $2.5\times10^{-10}$ \cite{Electron}, one can put constraint
on the value of $\lambda_0$, so that it still will be valid to the measurement of $\alpha^{-1}$ in the presence of
QG effects. This is known as a bottom-up QG phenomenological approach in which our analysis
in this study is based on it. Indeed, an interesting idea transposed into this approach
is that relying on data due to low energy phenomena, there is the possibility of
probing Planck scale spacetime with different resolutions depending on the relevant precision
measurements. Now by setting values of $R_\infty=10973731.565839(55) m^{-1},\,m_{Rb}
=86.909180535(10) m_u,\, m_e=5.4857990946(22)\times 10^{-4} m_u,\, M_{P}\sim1.3\times10^{19}m_u,
\,c=3\times10^8$m/s, the following results are obtained for bounds on $\lambda_0$
\begin{equation}\label{e-A5a}
-2<\lambda_0<-1
\end{equation}
and
\begin{equation}\label{e-A5b}
1<\lambda_0<2\,,
\end{equation}
for two cases of positive and negative signs, respectively. As we have mentioned previously,
the condition $\lambda_0>0$ should be satisfied. So, the first constraint is rejected.\\

By setting $n=1$,  Eq. (\ref{e-3}) in one dimension reads off as
\begin{eqnarray}\label{e-A6}
 p  \to \tilde p =  p (1 \pm \lambda_1p),~~~~~~~\lambda_2=\frac{\lambda_0}{(M_pc)}~.
\end{eqnarray}
In this new representation, by keeping all powers of $\lambda_0$
the relevant GUP modified kinetic energy of $^{87}$Rb atoms can also be written as
\begin{equation}\label{e-A7}
E_{K-GUP(n=1)}=\frac{p^2}{2 m_{Rb}}+\lambda_0^2\frac{p^4}{2 m_{Rb}(M_pc)^2}\pm
\lambda_0\frac{p^3}{m_{Rb}(M_pc)}~.
\end{equation}
Here the GUP counterpart of the Eq. (\ref{select-means}) takes the following form
\begin{equation}\label{e-A8}
|\delta_{sel}-\delta_{meas}|_{GUP(n=1)}= (k_1+k_2)|u_f-u_i|+\frac{2m_{Rb}^{2}\lambda_0^2(k_1+k_2)}{(M_pc)^2}
(u_{f}^{3}-u_{i}^{3})\pm\frac{3m_{Rb}\lambda_0(k_1+k_2)}{(M_pc)}(u_{f}^{2}-u_{i}^{2})~,
\end{equation} where by putting it into Eqs. (\ref{h/m}) and (\ref{alpha}) finally we
will arrive at
\begin{equation}\label{e-A9}
\alpha^{-1}_{GUP(n=1)}=\bigg[\frac{2R_\infty}{c}\,\frac{m_u}{m_e}\,\frac{h}{m_Rb}\bigg(1+\frac{2(\frac{m_{Rb}}{m_u})^{2}\lambda_0^2
}{(\frac{M_p}{M_u}c)^2}(u_{f}^{2}+u_iu_f+u_{i}^{2})\pm\frac{3\frac{m_{Rb}}{m_u}\lambda_0}{(\frac{M_p}{M_u}c)}(u_{f}+u_{i})\bigg)
\bigg]^{-\frac{1}{2}}\,.
\end{equation}
Using the same method described as above, we obtain the following upper bounds
on $\lambda_0$ as
\begin{equation}\label{e-A9a}
 1 \leq \lambda_0 < 1.5\times10^{14}
\end{equation}
and
\begin{equation}\label{e-A9b}
1\leq\lambda_0<4\times10^{23}
\end{equation}
for two cases of positive and negative signs, respectively. \\

By setting \textbf{$n=2$},  Eq. (\ref{e-3}) in one dimension
reads off as
\begin{eqnarray}\label{e-A10}
 p  \to \tilde p =  p (1 \pm \lambda_2p^2),~~~~~~~\lambda_2=\frac{\lambda_0}{(M_pc)^2}~.
\end{eqnarray}
Here also by keeping all powers of $\lambda_0$ , the relevant
GUP modified kinetic energy of $^{87}$Rb atoms is presented as follows
\begin{equation}\label{e-A11}
E_{K-GUP(n=2)}=\frac{p^2}{2 m_{Rb}}+\lambda_0^2\frac{p^6}{2 m_{Rb}(M_pc)^4}\pm
\lambda_0\frac{p^4}{m_{Rb}(M_pc)^2}
\end{equation} where results in the following modified expression
\begin{equation}\label{e-A12}
|\delta_{sel}-\delta_{meas}|_{GUP(n=2)}= (k_1+k_2)|u_f-u_i|+\frac{3m_{Rb}^{4}\lambda_0^2(k_1+k_2)}{(M_pc)^4}
(u_{f}^{5}-u_{i}^{5})\pm\frac{4m_{Rb}\lambda_0(k_1+k_2)}{(M_pc)^2}(u_{f}^{3}-u_{i}^{3})~,
\end{equation} for Eq. (\ref{select-means}). Putting it into Eqs. (\ref{h/m}) and (\ref{alpha})
 we arrive at
\begin{equation}\label{e-A13}
\alpha^{-1}_{GUP(n=2)}=\bigg[\frac{2R_\infty}{c}\,\frac{m_u}{m_e}\, \frac{h}{m_{Rb}}\bigg(1+\frac{3(\frac{m_{Rb}}{m_u})^{4}\lambda_0^2
}{(\frac{M_p}{M_u}c)^4} \,\frac{u_{f}^{5}-u_{i}^{5}}{u_{f}-u_{i}}\pm\frac{4(\frac{m_{Rb}}{m_u})^2\lambda_0}{(\frac{M_p}{M_u}c)^2}
(u_{f}^{2}+u_iu_f+u_{i}^{2})\bigg)\bigg]^{-\frac{1}{2}}
\end{equation}
Using the method described previously, we obtain the following upper bounds
on $\lambda_0$
\begin{equation}\label{e-A13a}
1\leq\lambda_0<1.4\times10^{38}~,
\end{equation}
and
\begin{equation}\label{e-A13b}
1\leq\lambda_0<3.2\times10^{47}~,
\end{equation}
for two cases of positive and negative sign, respectively. \\

By setting $n=3$,  Eq. (\ref{e-3}) in one dimension reads of as
\begin{eqnarray}\label{e-A14}
 p  \to \tilde p =  p (1 \pm \lambda_3p^3),~~~~~~\lambda_3=\frac{\lambda_0}{(M_pc)^3}~,
\end{eqnarray}
In the same way, for the relevant GUP modified kinetic energy of
$^{87}$Rb atoms, we have
\begin{equation}\label{e-A15}
E_{K-GUP(n=3)}=\frac{p^2}{m_{Rb}}+\lambda_0^2\frac{p^8}{2 m_{Rb}(M_pc)^6}\pm
\lambda_0\frac{p^5}{m_{Rb}(M_pc)^3}~.
\end{equation}
In the presence of the relevant GUP modification, Eq. (\ref{select-means}) is rewritten as
\begin{equation}\label{e-A16}
|\delta_{sel}-\delta_{meas}|_{GUP(n=3)}= (k_1+k_2)|u_f-u_i|+\frac{8m_{Rb}^{6}\lambda_0^2(k_1+k_2)}{2(M_pc)^6}
(u_{f}^{7}-u_{i}^{7})\pm\frac{5m_{Rb}\lambda_0(k_1+k_2)}{(M_pc)^3}(u_{f}^{5}-u_{i}^{5})~,
\end{equation} where by putting it into Eqs. (\ref{h/m}) and (\ref{alpha}), we come to
\begin{equation}\label{e-A17}
\alpha^{-1}_{GUP(n=3)}=\bigg[\frac{2R_\infty}{c} \, \frac{m_u}{m_e} \, \frac{h}{m_{Rb}}\bigg(1+\frac{4(\frac{m_{Rb}}{m_u})^{6}\lambda_0^2
}{(\frac{M_p}{M_u}c)^6} \, \frac{u_{f}^{7}-u_{i}^{7}}{u_{f}-u_{i}}\pm\frac{5(\frac{m_{Rb}}{m_u})^3\lambda_0}{(\frac{M_p}{M_u}c)^3}
(u_{f}^{3}+u_i^{2}u_f+ u_iu_f^2+u_{i}^{3})\bigg)\bigg]^{-\frac{1}{2}}\,.
\end{equation}
Once again via the method described above, we obtain other upper bounds on $\lambda_0$
as follows
\begin{equation}\label{e-A17a}
1\leq\lambda_0<1.5\times10^{62}~,
\end{equation}
and
\begin{equation}\label{e-A17b}
1\leq\lambda_0<3.3\times10^{71}~,
\end{equation} for two cases of positive and negative signs, respectively. \\

Comparing these obtained constraints for different values of $n$ shows obviously that by considering higher order powers of momentum
in the deformed coordinate representation and neglecting all smaller order terms in momentum in each step, the possibility
for accurate resolution of spacetime structure in Planck scale reduces for higher order terms.

\subsection{The case with superposition of the GUP modified terms}

Here, unlike the previous subsection, we treat a general form of GUP
as a superposition of all possible modification terms (we restrict ourselves to $n=0,1,2,3$).
In this regard, for the one-dimensional system Eq. (\ref{e-3}) takes the following form
\begin{eqnarray}\label{e-B0}
p  \to \tilde p =  p \left(1 \pm \lambda_0\pm\lambda_1 p\pm \lambda_2 p^2
\pm\lambda_3 p^3\pm....  \right)~.
\end{eqnarray}
In this case, involving quantities $E_{K},~|\delta_{sel}-\delta_{meas}|$
and $\alpha^{-1}$ can be modified in the following forms
\begin{eqnarray}\label{e-B1}
E_{K-GUP(\pm)}=\frac{1}{2m_{Rb}}\bigg[(1\pm\lambda_0)^2p^2+
(2\lambda_0\lambda_1\pm2\lambda_1)p^3+(\lambda_{1}^{2}+2\lambda_0\lambda_2\pm2\lambda_2)p^4 \\ \nonumber+
(2\lambda_0\lambda_3+2\lambda_1\lambda_2\pm2\lambda_{3})p^5+(\lambda_2^2+2\lambda_1\lambda_3)p^6
+2\lambda_2\lambda_3p^7+\lambda_3^{2}p^8\bigg]\,,
\end{eqnarray}

\begin{eqnarray}\label{e-B3}
&|\delta_{sel}-\delta_{meas}|_{GUP(\pm)}=(1\pm\lambda_0)^2(k_1+k_2)|u_f-u_i|+6m_{Rb}(\lambda_0\lambda_1\pm\lambda_1)
(k_1+k_2)(u_{f}^{2}-u_{i}^{2})\nonumber\\
&+2m_{Rb}^{2}(\lambda_{1}^{2}+2\lambda_0\lambda_2\pm2\lambda_2)(k_1+k_2)(u_{f}^{3}-u_{i}^{3})
+5m_{Rb}^{3}(\lambda_0\lambda_3+\lambda_1\lambda_2\pm\lambda_{3})(k_1+k_2)(u_{f}^{4}-u_{i}^{4})\nonumber\\
&+9m_{Rb}^{4}(2\lambda_1\lambda_3+\lambda_2^2)(k_1+k_2)(u_{f}^{5}-u_{i}^{5})+14m_{Rb}^{5}\lambda_2\lambda_3(k_1+k_2)(u_{f}^{6}-u_{i}^{6})
4m_{Rb}^{6}\lambda_2\lambda_3(k_1+k_2)(u_{f}^{7}-u_{i}^{7})&~,
\end{eqnarray}
and
\begin{eqnarray}\label{e-B4}
&\alpha^{-1}_{GUP(\pm)}=\bigg[\frac{2R_\infty}{c}\,\frac{m_u}{m_e}\,\frac{h}{m_{Rb}}\bigg(1+(1\pm\lambda_0)^2+6\frac{m_{Rb}}
{m_u}(\lambda_0\lambda_1\pm\lambda_1)(u_{f}+u_{i})+2(\frac{m_{Rb}}{m_u})^{2}(\lambda_{1}^{2}+2\lambda_0\lambda_2\pm2\lambda_2)
(u_{f}^{2}+u_iu_f+u_{i}^{2})\nonumber\\
&+5(\frac{m_{Rb}}{m_u})^{3}(\lambda_0\lambda_3+\lambda_1\lambda_2\pm\lambda_{3})(\frac{u_{f}^{4}-u_{i}^{4}}{u_f-u_i})
+9(\frac{m_{Rb}}{m_u})^{4}(\lambda_2^2+2\lambda_1\lambda_3)(\frac{u_{f}^{5}-u_{i}^{5}}{u_f-u_i})\nonumber\\
&+14(\frac{m_{Rb}}{m_u})^{5}\lambda_2\lambda_3(\frac{u_{f}^{6}-u_{i}^{6}}{u_f-u_i})+
4(\frac{m_{Rb}}{m_u})^{6}\lambda_2\lambda_3(\frac{u_{f}^{7}-u_{i}^{7}}{u_f-u_i})\bigg)\bigg]^{-\frac{1}{2}}&~,
\end{eqnarray} respectively. By setting $\lambda_2=\lambda_1^2$ and $\lambda_3=\lambda_1^3$,
the following upper bounds on $\lambda_0$ are achieved
\begin{equation}\label{e-B5}
-1.00002<\lambda_0<-0.999978
\end{equation}
and
\begin{equation}\label{e-B6}
0.999978<\lambda_0<1.00002
\end{equation} for the cases of positive and negative signs, respectively.
The case with negative $\lambda_{0}$ is not acceptable as we have discussed previously.
The condition (\ref{e-B6}) is very tight bound on quantum gravity parameter, $\lambda_0$.
So, in this proposed setup one is able to probe the quantum spacetime
structure even in low energy regime. This is so important in testing QG
proposal in laboratory in energy scales accessible today.\\

To summarize, in this work, via an accessible low energy technology,
we found possibility of probing the quantum spacetime structure with a
favorable theoretical resolution. Concerning the GUP proposal containing
the most general modification of coordinate representation
of the momentum operator, we have studied how to use the outcomes from $^{87}$Rb CARE (specially the
measurement of FSC) schematically designed in Figs. \ref{Fig1} and \ref{Fig2} to constrain
the relevant dimensionless, QG characteristic parameter $\lambda_0$. By enclosure the underlying
AI setup within an high finesse optical cavity, which is
supported by a new class of low power consumption integrated
devices known as MEOMS, it has been distinguished from
the conventional architectures. MEOMS belong to the family
of microsystems technology which along with electro mechanical
energy, harness the low dimension optics. As a distinct factor in this
setup, it should be pointed out the presence of a micro mechanical
oscillator instead of spherical confocal mirrors as one of the
components of high finesse optical cavity. \\

Altogether, this study has been done in two fashion: with and without superposition of separate modification terms in momentum.
The first step was based on the assumption that there
is no superposition of modification terms
so that each term is considered individually
in the absence of other possible terms. In this way,
the best constraint on $\lambda_0$ is extracted
as $1\leq\lambda_0<2$, when in the deformed coordinate
representation, makes the momentum operator to appear as a
re-scaled quantity. However, the constraints explicitly
reflect the fact that by increasing the power of
momentum in deformed coordinate representation
the resolution for exploring the spacetime structure in Planck's scale reduces and becomes much weaker.\\
Contrary to the first step, in the next step we have considered a general form of GUP
as a superposition of all possible modification terms (we preferred to consider just four first terms with $n=0,1,2,3$). We have obtained
a strict and tight constraint as $0.999978<\lambda_0<1.00002$
which is very favorable from the theoretical perspective.
This constraint obviously challenges the long-lived belief in QG phenomenology project that probing
quantum spacetime structure is possible just by having full access
to Planck energy. In other words, this setup paves a novel way for probing Planck scale spacetime procured by
the accessible low energy technology, in particular the
proposed opto-atomic $^{87}$Rb interferometric arrangement. So, one may argue that the best playground
to test QG is not just the high energy physics; a table-top nanosystem assembly can act as well.\\

%{\bf Additional Information}\\
%There is no Competing Financial Interests for this work.
\section{Acknowledgments}
The work of K. Nozari has been financially supported by
Research Institute for Astronomy and Astrophysics of
Maragha (RIAAM) under research project No. 1/5440-67.


\begin{thebibliography}{}
\bibitem{H2003}
S. Hossenfeldera et al., Collider signatures in the Planck regime, Phys. Lett. B \textbf{575} (2003) 85
\bibitem{Hedrich2009}
R. Hedrich, Quantum Gravity: Has Spacetime Quantum Properties?,	[arXiv:0902.0190 [gr-qc]]
\bibitem{st1}
D. J. Gross and P. F. Mendle, String Theory Beyond the Planck Scale, Nucl. Phys. B {\bf 303} (1988) 407
\bibitem{st2}
K. Konishi, G. Paffuti and P. Provero, Minimum Physical Length and the Generalized Uncertainty Principle in String Theory,
 Phys. Lett. B {\bf 234} (1990) 276
\bibitem{Lee}
C. Rovelli and L. Smolin, Discreteness of area and volume in quantum gravity, Nucl. Phys. B \textbf{442} (1995) 593
\bibitem{Ashtekar}
A. Ashtekar and J. Lewandowski, Quantum theory of geometry. 1: Area operators, Class. Quantum. Grav. \textbf{14} (1997) A55
\bibitem{dsr1}
G. Amelino-Camelia, Doubly special relativity: First results and key open problems, Int. J. Mod. Phys. D \textbf{11} (2002) 1643
\bibitem{dsr2}
J. Magueijo and L. Smolin, Lorentz invariance with an invariant energy scale, Phys. Rev. Lett.  \textbf{88} (2002) 190403
\bibitem{Ali2009}
A. F. Ali, S. Das and E. C. Vagenas, Discreteness of Space from the Generalized Uncertainty Principle, Phys. Lett. B \textbf{678} (2009) 497
\bibitem{Das2010}
S. Das, E. C. Vagenas and A. F. Ali, Discreteness of Space from GUP II: Relativistic Wave Equations, Phys. Lett. B \textbf{690} (2010) 407
\bibitem{Kempf1}
A. Kempf, G. Mangano, R. B. Mann, Hilbert space representation of the minimal length uncertainty relation, Phys. Rev. D \textbf{52} (1995) 1108
\bibitem{Sabin2012}
S. Hossenfelder, Minimal Length Scale Scenarios for Quantum Gravity, Living Rev. Rel \textbf{16} (2013) 2
\bibitem{Kempf2}
A. Kempf, G. Mangano, Minimal length uncertainty relation and ultraviolet regularization, Phys. Rev. D \textbf{55} (1997) 7909
\bibitem{Kempf3}
A. Kempf, Nonpointlike particles in harmonic oscillators, J. Phys. A \textbf{30} (1997) 2093
\bibitem{Amati}
D. Amati, M. Ciafaloni, G. Veneziano, Can Space-Time Be Probed Below the String Size?, Phys. Lett. B \textbf{216} (1989) 41
\bibitem{M1}
M. Maggiore, A Generalized uncertainty principle in quantum gravity, Phys. Lett. B \textbf{304} (1993) 65
\bibitem{M2}
M. Maggiore, The Algebraic structure of the generalized uncertainty principle, Phys. Lett. B \textbf{319} (1993) 83
\bibitem{M3}
M. Maggiore, Quantum groups, gravity and the generalized uncertainty principle, Phys. Rev. D \textbf{49} (1994) 5182
\bibitem{Adler2001}
Ronald J. Adler, Pisin Chen, David I. Santiago, The Generalized Uncertainty Principle and Black Hole Remnants, Gen. Rel. Grav. \textbf{33} (2001) 2101
\bibitem{Chen2014}
P. Chen, Y. Chin Ong, D. Yeom, Generalized Uncertainty Principle: Implications for Black Hole Complementarity, JHEP \textbf{12} (2014) 021
\bibitem{Black2012}
S. Das and A. F. Ali, No Existence of Black Holes at LHC Due to Minimal Length in Quantum Gravity, JHEP \textbf{1209} (2012) 067
\bibitem{CMS2011}
\textbf{CMS Collaboration Collaboration}, V. Khachatryan et al., Search for Microscopic Black Hole Signatures at the Large Hadron Collider,
Phys. Lett. B \textbf{697} (2011) 434
\bibitem{Sabri}
Y. Sabri and K. Nouicer, Phase transitions of a GUP-corrected Schwarzschild black hole within isothermal cavities,
Class. Quant. Grav. \textbf{29} (2012) 215015
\bibitem{ref:PLB}
M. Khodadi, K. Nozari, A. Hajizadeh, Some Astrophysical Aspects of a Schwarzschild Geometry Equipped with a Minimal Measurable Length, Phys. Lett. B \textbf{770} (2017) 556
\bibitem{ref:JCAP}
N. Khosravi and H. R. Sepangi, The Cosmological implications of a fundamental length: A DSR inspired de-Sitter spacetime, JCAP \textbf{0804} (2008) 011
\bibitem{kh2008}
N. Khosravi and H. R. Sepangi, A Fundamental length as a candidate for dark energy: A DSR inspired FRW spacetime, Phys. Lett. A \textbf{372} (2008) 3356
\bibitem{Tawfik}
A. Tawfik, H. Magdy and A. Farag Ali, Effects of quantum gravity on the inflationary parameters and thermodynamics of the early universe, Gen. Rel. Grav. \textbf{45} (2013) 1227
\bibitem{ref:PLB2016}
M. A. Gorji, Late time cosmic acceleration from natural infrared cutoff, Phys. Lett. B \textbf{760} (2016) 769
\bibitem{ref:GRG}
M. Khodadi, K. Nozari and H. R. Sepangi, More on the initial singularity problem in gravity's rainbow cosmology, Gen. Rel. Grav. \textbf{48} (2016) 166
\bibitem{ref:Ann}
M. Khodadi, K. Nozari, Some Features of Scattering Problem in a $\kappa$-Deformed Minkowski Spacetime, Annalen der Physik \textbf{528} (2016) 785
\bibitem{ref:CQG}
M. Khodadi, K. Nozari, E. N. Saridakis, Emergent universe in theories with natural UV cutoffs, Class. Quant. Grav. \textbf{35} (2018) 015010
\bibitem{Reza2016}
R. Rashidi, Generalized Uncertainty Principle Removes The Chandrasekhar Limit, Annals Phys. \textbf{374} (2016) 434
\bibitem{Yen2018}
Y. C. Ong, Generalized Uncertainty Principle, Black Holes, and White Dwarfs: A Tale of Two Infinities, JCAP \textbf{09} (2018) 015

\bibitem{Kumar2018}
D. K. Mishra, N. Chandra, Quantum gravity corrections to white dwarf dynamics, [arXiv:1803.06640 [gr-qc]]
\bibitem{WEP1}
V. M. Tkachuk, Deformed Heisenberg algebra with minimal length and equivalence principle, Phys. Rev. A \textbf{86} (2012) 062112
\bibitem{WEP2}
S. Ghosh, Quantum Gravity Effects in Geodesic Motion and Predictions of Equivalence Principle Violation, Class. Quant. Grav. \textbf{31} (2014) 025025
\bibitem{Interf2015}
L. Zhou et al., Test of Equivalence Principle at $10^{-8}$ Level by a Dual-species Double-diffraction Raman Atom Interferometer,
Phys. Rev. Lett. \textbf{115} (2015) 013004
\bibitem{Amelino}
G. Amelino-Camelia, Quantum-Spacetime Phenomenology, Living Rev. Rel. \textbf{16} (2013) 5
\bibitem{Sabin}
S. Hossenfelder, Experimental Search for Quantum Gravity, [arXiv:1010.3420 [gr-qc]]
\bibitem{Stachel}
J. Stachel, \emph{``Early history of quantum gravity, Black Holes, Gravitational Radiation and the Universe''},
edited by B. Iyer and B. Bhawal, Kluwer Academic Publisher, Netherlands, (1999)
\bibitem{Lect2000}
G. Amelino-Camelia, Are we at the dawn of quantum gravity phenomenology? Lect. Notes Phys. \textbf{541} (2000) 1
\bibitem{G1998}
G. Amelino-Camelia et al., Tests of quantum gravity from observations of gamma-ray bursts, Nature \textbf{393} (1998) 763
\bibitem{DV1998}
D. V. Ahluwalia, Can general relativistic description of gravitation be considered complete? Mod. Phys. Lett. A \textbf{13} (1998) 1393
\bibitem{G1999}
G. Amelino-Camelia, An Interferometric gravitational wave detector as a quantum gravity apparatus, Nature \textbf{398} (1999) 216
\bibitem{GF1999}
G. F. Giudice, R. Rattazzi and J. D. Wells, Quantum gravity and extra dimensions at high-energy colliders, Nucl. Phys. B \textbf{544} (1999) 3
\bibitem{UP0}
S. Das and E. C. Vagenas, Universality of Quantum Gravity Corrections, Phys. Rev. Lett. \textbf{101} (2008) 221301
\bibitem{UP1}
P. Pedram, K. Nozari and S. H. Taheri, The effects of minimal length and maximal momentum on the transition rate of ultra cold neutrons in gravitational field, JHEP \textbf{1103} (2011) 093
\bibitem{UP2}
A. F. Ali, S. Das and E. C. Vagenas, A proposal for testing Quantum Gravity in the lab, Phys. Rev. D {\bf 84} (2011) 044013
\bibitem{UP3}
P. Wang, H. Yang and X. Zhang, Quantum gravity effects on compact star cores, Phys. Lett. B \textbf{718} (2012) 265
\bibitem{UP4}
E. Castellanos and G. Chacon-Acosta, Polymer Bose–Einstein condensates, Phys. Lett. B \textbf{722} (2013) 119
\bibitem{UP5}
S. Ghosh, Quantum Gravity Effects in Geodesic Motion and Predictions of Equivalence Principle Violation, Class. Quantum Grav. \textbf{31}  (2014) 025025
\bibitem{UP6}
S. Jalalzadeh, M. A. Gorji and K. Nozari, Deviation from the Standard Uncertainty Principle and the Dark Energy Problem,
Gen. Rel. Grav. \textbf{46} (2014) 1632
\bibitem{UP7}
K. Nozari, M. Khodadi, M. A. Gorji, Bounds on quantum gravity parameter from the SU(2) NJL effective model of QCD,
Europhys. Lett. \textbf{112} (2015) 60003
\bibitem{UP8}
G. Chacon-Acosta and H. Hernandez-Hernandez, Polymer quantum effects on compact stars models,
Int. J. Mod. Phys. D \textbf{24} (2015) 1550033
\bibitem{Nature}
I. Pikovski et al., Probing Planck-scale physics with quantum optics, Nature Phys. \textbf{8} (2012) 393
\bibitem{NPB}
S. Dey et al., Probing noncommutative theories with quantum optical experiments, Nucl. Phys. B \textbf{924} (2017) 578
\bibitem{Sci}
M. Khodadi et al., A new bound on polymer quantization via an opto-mechanical setup, Sci. Rep. \textbf{8} (2018) 1659
\bibitem{Accel}
A. Peters, K. Y. Chung, and S. Chu, Measurement of gravitational acceleration by dropping atoms, Nature \textbf{400} (1999) 849
\bibitem{Gyro}
T. L. Gustavson, A. Landragin, and M. A. Kasevich, Rotation sensing with a dual atom-interferometer Sagnac
gyroscope, Class. Quantum Grav. \textbf{17} (2000) 2385
\bibitem{Map}
C. Jentsch et al., High precision atom interferometry on ground and space, Gen. Rel. Grav. \textbf{36} (2004) 2197
\bibitem{Setup2006}
P. Clade et al., Precise measurement of $\hbar/m_{Rb}$ using Bloch oscillations in a vertical optical lattice: Determination of the fine-structure constant, Phys. Rev. A \textbf{74 } (2006) 052109
\bibitem{PRA2016}
D. Gao, M. Zhan, Constraining the generalized uncertainty principle with cold atoms, Phys. Rev. A \textbf{94} (2016) 013607
\bibitem{RMP2009}
A. D. Cronin, J. Schmiedmayer, and D. E. Pritchard, Optics and interferometry with atoms and molecules,
Rev. Mod. Phys. \textbf{81} 1051 (2009)
\bibitem{PRL2018}
R. Geiger, M. Trupke, Proposal for a Quantum Test of the Weak Equivalence Principle with Entangled Atomic Species, Phys. Rev. Lett. \textbf{120} (2018) 043602
\bibitem{Atomic2002}
A. Wicht et al., A Preliminary Measurement of the Fine Structure Constant Based on Atom Interferometry,
Phys. Scr. \textbf{102} (2002) 82
\bibitem{Camelia2009}
G. Amelino-Camelia et al., Constraining the Energy-Momentum Dispersion Relation with Planck-Scale Sensitivity Using Cold Atoms,
 Phys. Rev. Lett. \textbf{103} (2009) 171302
\bibitem{Alpha}
R. Bouchendira et al., New determination of the fine structure constant and test of the quantum electrodynamics,
Phys. Rev. Lett. \textbf{106} (2011) 080801
\bibitem{Electron}
T. Aoyama et al., Tenth-Order QED Contribution to the Electron g-2 and an Improved Value of the Fine Structure Constant
 Phys. Rev. Lett. \textbf{109} (2012) 111807
\bibitem{Mir2016}
S. Masood et al., The most general form of deformation of the Heisenberg algebra from the generalized uncertainty principle,
Phys. Lett. B \textbf{763} (2016) 218
\bibitem{Ye2018}
Y. C. Ong, Yuan Yao, Generalized Uncertainty Principle and White Dwarfs Redux: How Cosmological Constant Protects Chandrasekhar Limit,
Phys. Rev. D \textbf{98} (2018) 126018
\bibitem{Jizba2010}
P. Jizba, H. Kleinert and F. Scardigli, Uncertainty Relation on World Crystal and its Applications to Micro Black Holes, Phys. Rev. D \textbf{81} (2010) 084030
\bibitem{G2012}
G. Kioseoglou et al., Valley polarization and intervalley scattering in monolayer $MoS_{2}$,
Appl. Phys. Lett. \textbf{101} (2012) 221907
\bibitem{Muller2009}
T. M\"{u}ller et al., A compact dual atom interferometer gyroscope based on laser-cooled rubidium,
Eur. Phys. J. D \textbf{53} (2009) 273



\end{thebibliography}
\end{document}